# A Monte Carlo investigation of secondary electron emission from solid targets: spherical symmetry versus momentum conservation within the classical binary collision model

Maurizio Dapor

*FBK-Center for Materials and Microsystems, via Sommarive 18, I-38050 Povo, Trento, Italy*

**Abstract**

A Monte Carlo scheme is described where the secondary electron generation has been incorporated. The initial position of a secondary electron due to Fermi sea excitation is assumed to be where the inelastic collision took place, while the polar and azimuth angles of secondary electrons can be calculated in two different ways. The first one assumes a random direction of the secondary electrons, corresponding to the idea that slow secondary electrons should be generated with spherical symmetry. Such an approach violates momentum conservation. The second way of calculating the polar and azimuth angles of the secondary electrons takes into account the momentum conservation rules within the classical binary collision model. The aim of this paper is to compare the results of these two different approaches for the determination of the energy distribution of the secondary electrons emitted by solid targets.

## 1. Introduction

A Monte Carlo scheme is briefly described and utilized to follow the entire cascade of the secondary electrons generated by an incident electron beam in silicon and copper. The Monte Carlo code takes into account the main inelastic and elastic interactions of electrons in solid targets.

The main inelastic interactions of an electron in a solid concern the production of electron transitions from the valence to the conduction band and the collective oscillations generation (plasmons). If the electron energy is high enough, inelastic collisions with inner-shell electrons and consequent ionizations can occur as well. The elastic collisions with the screened potentials of the atoms are, on the other hand, mainly responsible of the electron deflections.

Secondary electrons due to the inelastic interactions generate further secondary electrons so that a cascade process occurs.

Monte Carlo simulated results of the secondary electron spectra are first obtained assuming that the initial angles of emission of the secondary electrons have random directions. The spectra obtained in such a way are then compared to the same spectra calculated assuming that the initial angles of emission of the secondary electrons inside the solid satisfy conditions obtained assuming momentum conservation within the limits of the classical binary collision model.

## 2. The Monte Carlo scheme

The Monte Carlo scheme has been described elsewhere [1], so that we limit ourselves to briefly summarize the main topics which are specific of the secondary electron emission.

Comparison with experimental data concerning energy losses of primary electrons can be found in [2-5], while comparison with experimental data about secondary electrons has been presented in [6].

The differential elastic scattering cross sections have been calculated by the relativistic partial wave expansion method [7-10] while the differential inelastic scattering cross sections have been computed by using the Ritchie theory [11]. Within the Drude-Lorentz model, the dielectric function $\varepsilon(\omega,k)$ is given by a superposition of free and bound oscillators,

$$\varepsilon(\omega,k) = 1 - \omega_p^2 \sum_n \frac{f_n}{\omega^2 - \omega_n^2 - \omega_k^2 + i\omega\Gamma_n}, \quad (1)$$

where $\omega_p$ is the plasma frequency, $\omega_n$ are the characteristic excitation frequencies, $\Gamma_n$ are positive damping coefficients which can depend on transferred momentum $k$ but do not depend on the frequency $\omega$ and $f_n$ are the fractions of the valence electrons bound with energies $\hbar\omega_n$. $\hbar\omega_k$ is an energy, related to the dispersion relation, that, for high values of the transferred momentum $k$, approaches a free particle form [12]. According to Yubero and Tougaard [13] and Cohen Simonsen et al. [14] the simple relation $\hbar\omega_k = \hbar^2 k^2/2m$, where $m$ is the electron mass, has been used in this work.

The particles are followed until they emerge from the surface or until their energy become lower than the potential barrier between the vacuum level and the minimum of the conduction band. Each simulation was obtained by calculating all the electron trajectories produced by $10^6$ primary electrons. The whole cascade of the secondary electrons has been followed.

## 3. Transmission coefficient

When a secondary electron reaches the target surface, it can be emitted only if its energy, $E$, and direction with respect to the normal to the surface, $\vartheta_1$, satisfy the condition

$$E\cos^2\vartheta_1 \geq \chi, \qquad (2)$$

where $\chi$ is the potential barrier between the vacuum level and the minimum of the conduction band.

The quantum mechanical transmission coefficient was utilized in the present calculation. It is computed on the basis of the following considerations. Let us consider, along the $z$ direction, two regions, respectively, inside and outside the solid target and, at $z=0$, a potential barrier $\chi$. The first region (inside the solid) corresponds to the following solution of the Schrödinger equation

$$\varphi_1 = A_1 \exp(ik_1 z) + B_1 \exp(-ik_1 z). \qquad (3)$$

The solution of the Schrödinger equation in the second region (the vacuum) is

$$\varphi_2 = A_2 \exp(ik_2 z). \qquad (4)$$

In these equations $k_1$ is the electron momentum in the solid

$$k_1 = \sqrt{\frac{2mE}{\hbar^2}} \cos\vartheta_1 \qquad (5)$$

and $k_2$ is the electron momentum in the vacuum

$$k_2 = \sqrt{\frac{2m(E-\chi)}{\hbar^2}} \cos\vartheta_2. \qquad (6)$$

$A_1$, $B_1$, and $A_2$ are constants. Notice that $\vartheta_2$ is the angle of emergence of the secondary electrons with respect to the normal to the surface *outside* the solid target, while $\vartheta_1$ is the same angle *inside* the solid target. The following conditions must be satisfied:

$$\varphi_1(0) = \varphi_2(0), \qquad (7)$$

$$\varphi_1'(0) = \varphi_2'(0), \qquad (8)$$

so that the transmission coefficient $T$ is given by

$$T = 1 - R = 1 - |B_1/A_1|^2 = \frac{4k_1 k_2}{(k_1 + k_2)^2}, \qquad (9)$$

where $R$ is the reflection coefficient. Taking into account of the definition of the electron momenta we get

$$T = \frac{4\sqrt{(1-\chi/E)\cos^2\vartheta_2/\cos^2\vartheta_1}}{\left[1+\sqrt{(1-\chi/E)\cos^2\vartheta_2/\cos^2\vartheta_1}\right]^2}. \qquad (10)$$

The conservation of the momentum parallel to surface imposes that

$$E\sin^2\vartheta_1 = (E-\chi)\sin^2\vartheta_2 \qquad (11)$$

and, as a consequence,

$$\cos^2\vartheta_1 = \frac{(E-\chi)\cos^2\vartheta_2 + \chi}{E}, \qquad (12)$$

$$\cos^2\vartheta_2 = \frac{E\cos^2\vartheta_1 - \chi}{E-\chi}, \qquad (13)$$

so that

$$T = \frac{4\sqrt{1-\chi/[(E-\chi)\cos^2\vartheta_2 + \chi]}}{\left[1+\sqrt{1-\chi/[(E-\chi)\cos^2\vartheta_2 + \chi]}\right]^2}, \qquad (14)$$

or, expressed as a function of $\vartheta_1$,

$$T = \frac{4\sqrt{1-\chi/(E\cos^2\vartheta_1)}}{\left[1+\sqrt{1-\chi/(E\cos^2\vartheta_1)}\right]^2}. \qquad (15)$$

Once calculated the transmission coefficient, which is given by Eq. (15) if the condition (2) is satisfied and zero otherwise, the code generates a random number, $\mu_1$, uniformly distributed in the range (0,1). It allows the secondary electron to be emitted into the vacuum if the condition $\mu_1 < T$ is satisfied. Otherwise, the secondary electron is specularly reflected without energy loss. Notice that the last part of the trajectory of the electrons which are not able to emerge (the specularly reflected ones) is followed by the code as well, as they can reach the surface again with the energy and angle necessary to emerge. Furthermore, during the last part of their travel, they can contribute to the entire cascade producing ulterior secondary electrons.

## 4. Initial polar and azimuth angle of the secondary electrons

The initial polar angle $\vartheta$ and the initial azimuth angle $\varphi$ of each secondary electron have been calculated in two different ways. In the first one, assuming that the secondary electrons emerge with spherical symmetry, their initial polar and azimuth angles are randomly determined as [15]

$$\vartheta = \pi \mu_2, \qquad (16)$$

$$\varphi = 2\pi \mu_3, \qquad (17)$$

where $\mu_2$ and $\mu_3$ are random numbers uniformly distributed in the range (0,1). Even if such an approach violates momentum conservation and it is therefore questionable, Shimizu and Ding noticed that, as slow secondary electrons are actually generated with spherical symmetry [15], it should be used and preferred when Fermi sea excitations are involved in the process of generation of secondary electrons. MC1 is the name attributed to the Monte Carlo code based on this method.

Since, to the author knowledge, a comparison with calculations where momentum conservation is taken into account was missing, a second code was utilized in which momentum conservation was taken into account by using the classical binary collision model so that, if $\theta$ and $\phi$ are, respectively, the polar and azimuth angle of the incident electron, then [15]

$$\sin \vartheta = \cos \theta, \qquad (18)$$

$$\varphi = \pi + \phi. \qquad (19)$$

We will indicate with MC2 the Monte Carlo code corresponding to this second approach.

## 5. Results and discussion

In Fig. 1 and in Fig. 2 the energy distributions of the secondary electrons emitted by silicon and copper targets, respectively, have been presented. The energy of the primary electron beam is $E_0=1000$ eV for the case of silicon, while it is $E_0=300$ eV for copper.

The Monte Carlo calculations obtained with the two different methods described above (MC1 and MC2) are compared to the Amelio theoretical results [16].

Using the Amelio's theory for the comparison, the MC1 scheme gives results that show a much better agreement than the MC2 code. Indeed, in the range of primary energies examined and for both the materials considered, it is clear that, using the MC1 code, the position of the maxima and the general trend of the energy distributions are in excellent agreement with the Amelio's data. On the other hand, the use of the MC2 code generates electron energy distributions which are not in so good agreement with the Amelio's data: the position of the maxima are shifted to higher energies and the shapes of the distributions are quite different from the Amelio's energy distributions. Notice that experimental data concerning secondary electron energy distributions were reported by Amelio as well. In Table 1 and in Table 2 the main features regarding the energy distributions [i.e. the Most Probable Energy (MPE) and the Half Width at Half Maximum (FWHM)] obtained with MC1 and MC2 are compared to the experimental data reported by Amelio.

Concerning the secondary electron yields, namely the measures of the areas under the energy distributions before performing normalization, they are summarized in Table 3 and in Table 4. The secondary electron yields both experimentally determined [17,18] and calculated with the MC1 code are considerably higher than the same quantities computed with the MC2 code. The comparison demonstrates as well that the MC1 code should be preferred to the MC2 code, in the primary energy range examined (300-1000 eV), because MC1 gives results in better agreement with the experimental ones than MC2.

The agreement of the MC1 results with the Amelio [16], Dionne [17] and Shimizu [18] theoretical and experimental data (and the disagreement between the MC2 and experimental results) can be attributed to the isotropy of the low-energy secondary electron emission due to (i) post-collisional effects and consequent random energy and momentum transfer among secondary electrons and (ii) to the interactions with the conduction electrons, just after secondary electrons are emitted, as well.

In conclusion, the results of the present investigation suggest that slow secondary electrons should be generated, in Monte Carlo codes, with spherical symmetry in order to get agreement to the available experimental and theoretical data.

## 6. Conclusion

An analysis of the results of two different Monte Carlo approaches for the determination of the energy distribution of the secondary electrons emitted by solid targets quantitatively confirms that, in Monte Carlo codes, slow secondary electrons should be generated with spherical symmetry in order to get agreement to the available experimental and theoretical data [15-18].

**Acknowledgments**

The author is indebted to Lucia Calliari (FBK, Trento) and Simone Taioli (FBK, Trento) for their stimulating comments and useful suggestions.

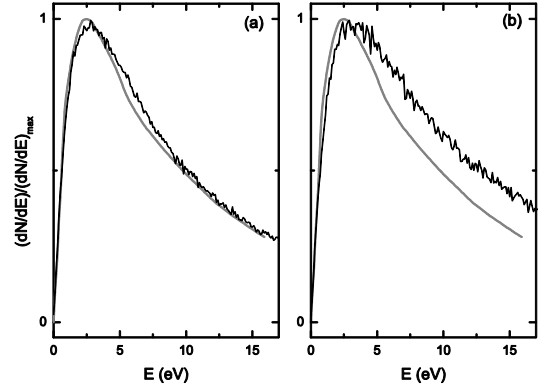

Fig. 2: Energy distribution of the secondary electrons emitted by a copper target. $E_0$=300 eV. The present Monte Carlo calculations (black lines) are compared to the Amelio theoretical results (gray lines) [16]. Panel (a): MC1code. Panel (b): MC2 code (see details in the text).

| Si(1.0 keV) | MC1 | MC2 | Experimental |
|---|---|---|---|
| MPE(eV) | 1.8 | 2.8 | 1.7 |
| FWHM(eV) | 5.3 | 8.5 | 5.0 |

Table 1: Monte Carlo Most Probable Energy (MPE) and Half Width at Half Maximum (FWHM) of secondary electron energy distributions obtained with two different schemes (MC1 and MC2: see details in the text). The experimental data were reported by Amelio [16]. Material on which the calculations and the measurements were carried out was bulks of Si irradiated by streams of electrons in the $+z$ direction. The primary energy of the incident electron beam was 1.0 keV.

| Cu(0.3 keV) | MC1 | MC2 | Experimental |
|---|---|---|---|
| MPE(eV) | 2.8 | 3.5 | 2.8 |
| FWHM(eV) | 9.2 | 12 | 10 |

Table 2: Monte Carlo Most Probable Energy (MPE) and Half Width at Half Maximum (FWHM) of secondary electron energy distributions obtained with two different schemes (MC1 and MC2: see details in the text). The experimental data were reported by Amelio [16]. Material on which the calculations and the measurements were carried out was bulks of Cu irradiated by streams of electrons in the $+z$ direction. The primary energy of the incident electron beam was 0.3 keV.

| $E_0$ (keV) | MC1 | MC2 | Dionne [17] |
|---|---|---|---|
| 0.3 | 1.26 | 0.58 | 1.17 |
| 0.5 | 1.15 | 0.54 | 1.12 |
| 1.0 | 0.91 | 0.46 | 0.94 |

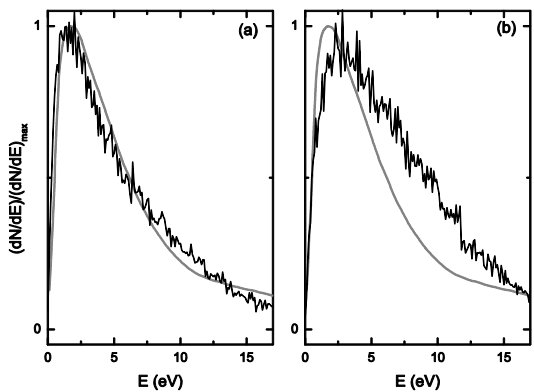

Fig. 1: Energy distribution of the secondary electrons emitted by a silicon target. $E_0$=1000 eV. The present Monte Carlo calculations (black lines) are compared to the Amelio theoretical results (gray lines) [16]. Panel (a): MC1code. Panel (b): MC2 code (see details in the text).

Table 3: Monte Carlo secondary electron yield calculations performed with two different schemes (MC1 and MC2: see details in the text). Material on which the calculations and the measurements were carried out was bulks of Si irradiated by streams of electrons in the +z direction. $E_0$ represents the primary energy of the incident electron beam.

| $E_0$ (keV) | MC1 | MC2 | Shimizu [18] |
|---|---|---|---|
| 0.3 | 1.09 | 0.71 | - |
| 0.5 | 1.02 | 0.65 | 1.01 |
| 1.0 | 0.81 | 0.53 | 0.89 |

Table 4: Monte Carlo secondary electron yield calculations performed with two different schemes (MC1 and MC2: see details in the text). Material on which the calculations and the measurements were carried out was bulks of Cu irradiated by streams of electrons in the +z direction. $E_0$ represents the primary energy of the incident electron beam.